\documentstyle[aps,epsf]{revtex}
\title{Instability of 1-loop superstring cosmology}
\author{Shinsuke {\sc Kawai}\footnote{E-mail:kawai@phys.h.kyoto-u.ac.jp}}
\address{Graduate School of Human and Environmental Studies, Kyoto University,
Kyoto 606-8501, Japan}
\author{Masa-aki {\sc Sakagami}\footnote{E-mail:sakagami@phys.h.kyoto-u.ac.jp}
 and Jiro {\sc Soda}\footnote{E-mail:jiro@phys.h.kyoto-u.ac.jp}}
\address{Department of Fundamental Sciences, FIHS, Kyoto University, Kyoto 
606-8501,
Japan}
\begin{document}
\maketitle
\begin{abstract}
A stability analysis is made in the context of the previously discovered 
non-singular cosmological solution from 1-loop corrected superstring 
effective action. 
We found that this solution has an instability in graviton mode, 
which is shown to have a close relation to the avoidance of initial singularity
via energy condition. We also estimate the condition for the 
breakdown of the background solution due to the overdominance of the 
graviton.

\end{abstract}

Tracing back the history of our presently expanding universe, we are naturally
led to the era of high temperature and large energy density. When the energy
density approaches Planck scale, our general relativistic description of the 
universe is no longer considered as valid 
since quantum gravitational processes are thought to be significant 
at such a high energy scale. The leading 
candidate for the satisfactory theory in the very early universe is the 
superstring theory\cite{sstr}, and roughly speaking, two approaches are being 
made in the attempt to construct the consistent history of the early universe. 

Pre-big-bang universe model\cite{pbb} was proposed to realize an inflation
\cite{inflation}``before'' the initial singularity.
Due to the scale factor duality\cite{sfduality}, the equations of motion in 
this model
have two distinct and disconnected branches of the solution, one corresponding
to an ordinary Friedmann solution, and the other corresponding to the 
so-called super-inflation solution with increasing Hubble parameter. The 
super-inflationary stage, which is usually put before the Friedmann
stage and characterized by the negative-power dependence of the scale factor on
time, is regarded as a sort of inflation that 
lessens the difficulties appearing in the standard big bang model. 
Although this model is advantageous in that the inflation is naturally 
explained from one of the most promising theories at present, 
many difficulties, besides the serious graceful-exit problem\cite{graceful},
have been pointed out\cite{ftuning,spectrum}.

Another approach in string cosmology is the search for non-singular 
cosmological models\cite{nonsingular}. Initial singularity problem is inherent
in the standard big bang cosmology\cite{initsing}, and unless we believe the 
universe was really generated by one sudden blow at infinite temperature, 
it is natural to think that the classical theory was violated in such an 
extreme
condition and that singular situation never happened. Also, getting rid of
the infinitely high temperature suggests possibilities to understand the 
whole history of our universe without knowing the theory of everything.

Recently, non-singular cosmological model based on 1-loop corrected superstring
effective action\cite{1loop} was proposed\cite{art94,maeda96}. 
One of the most interesting features of this model is that 
its solutions include a super-inflationary stage before the ordinary 
Friedmann-like universe, which is, just like pre-big-bang case, expected to 
give a natural explanation for the inflation.
It should be emphasized that the transition in this model is 
smooth in contrast with the pre-big-bang scenario.
In order to concentrate on the behavior of the metric-modulus sub-system
which plays the essential role in the realization of super-inflationary
singularity-free solution, a simpler version of this model was proposed by
Rizos and Tamvakis\cite{rt94}, 
in which they generalized the form of the coupling and analytically examined 
the conditions for the existence of non-singular solutions. 
In this letter perturbative analysis 
is made for this metric-modulus system and we show that its non-singular 
solution is geometrically unstable.


We start with the 1-loop effective action of the heterotic string
with orbifold compactification, which is given by\cite{tree,1loop,art94}:
\begin{equation}
{\cal S}=\int d^4x\sqrt{-g}\{\frac 12 R-\frac 14(D\Phi)^2-\frac 34 (D\sigma)^2
 +\frac 1{16} [\lambda_1 e^\Phi-\lambda_2\xi(\sigma)]R^2_{GB}\},
\label{eqn:ss1leAct}
\end{equation}
where $R$ ,$\Phi$ and $\sigma$ are the Ricci scalar curvature, the dilaton, 
and the modulus field, respectively\footnotemark.
\footnotetext{We will use in this paper the following convention.
 The signature of the metric: $(-,+,+,+),$ 
 Riemann tensor: $R^\mu{}_{\alpha\nu\beta}=\Gamma^\mu{}_{\alpha\beta,\nu}+...$,
 Ricci tensor: $R_{\alpha\beta}=R^\mu{}_{\alpha\mu\beta}$. Greek indices
 denote 4 dimensional space-time while Latin indices run from 1 to 3.}
The Gauss-Bonnet curvature is 
\begin{equation}
R^2_{GB}=R^{\mu\nu\kappa\lambda}R_{\mu\nu\kappa\lambda}
 -4 R^{\mu\nu}R_{\mu\nu}+R^2,
\end{equation}
and $\xi(\sigma)$ is an even function of the modulus field expressed with 
Dedekind $\eta$ function:
\begin{equation}
\xi(\sigma)=-\ln[2e^\sigma\eta^4 (i e^\sigma)]
 =-\ln 2-\sigma+\frac{\pi e^\sigma}{3}-4\sum_{n=1}^{\infty}\ln
(1-e^{-2n\pi e^\sigma}).
\label{eqn:ssxi}
\end{equation}
This action provides non-singular cosmological solutions driven by 
the modulus field for positive $\lambda_2$\cite{art94}.
These non-singular solutions experience a super-inflationary phase, 
which means the expansion of the universe with increasing Hubble parameter.
Since the non-singular nature of this model depends essentially on the behavior
of the modulus field\cite{art94}, the system is well represented by the one 
scalar field model\cite{rt94}:
\begin{equation}
{\cal S}=\int d^4x\sqrt{-g}\{\frac 12 R-\frac 12 (D\varphi)^2
 -\frac{\lambda}{16}\xi(\varphi) R_{GB}^2\}.
\label{eqn:rt94act}
\end{equation}
Variation of this action gives equations of motion
\begin{eqnarray}
G^\mu{}_\nu &=& \varphi^{;\mu}\varphi_{;\nu}
 -\frac 12 \varphi^{;\alpha}\varphi_{;\alpha}
 \delta^\mu{}_\nu  
 +\frac{\lambda}{2}\left[
   H^\mu{}_{\alpha\nu\beta}\xi^{;\alpha\beta}+G^\mu{}_\alpha
   \xi^{;\alpha}{}_{;\nu}-G^\mu{}_\nu\xi^{;\alpha}{}_{;\alpha}\right],
\label{eqn:Einstein}\\
\varphi^{;\alpha}{}_{;\alpha}&=&\frac{\lambda}{16}\xi_{,\varphi}R_{GB}^2.
\label{eqn:KleinGordon}
\end{eqnarray}
Here, ${}_{,\varphi}$ denotes derivative with respect to $\varphi$,
$G^\mu{}_\nu$ is the Einstein tensor and
\begin{equation}
H^\mu{}_{\alpha\nu\beta} := R^\mu{}_{\alpha\nu\beta}
-R_{\alpha\beta}\delta^\mu{}_\nu+R_{\alpha\nu}\delta^\mu{}_\beta.
\end{equation}
Assuming the homogeneous and isotropic flat metric 
\begin{eqnarray}
ds^2&=&-dt^2 + a^2 \delta_{ij}dx^i dx^j,
\label{eqn:bgmetric}
\end{eqnarray}
the equations of motion are rewritten using $\varphi$ and the Hubble parameter
$H:=\dot a/a$ as 
\begin{eqnarray}
&&\dot\varphi^2=6H^2 (1-\frac \lambda 2 H\dot\xi)
\label{eqn:bg1}\\
&&(2\dot H + 5H^2)(1-\frac \lambda 2 H \dot \xi)
 +H^2(1-\frac \lambda 2 \ddot\xi)=0
\label{eqn:bg2}\\
&&\ddot\varphi+3H\dot\varphi+\frac {3\lambda}{2}
 (\dot H+H^2)H^2\xi_{,\varphi}=0.
\label{eqn:bg3}
\end{eqnarray}
Dot ($\dot{}$) is the derivative with respect to the physical time $t$.
Thus, along with $\frac{d}{dt} a = aH$ and $\frac{d}{dt}\varphi = \dot\varphi$,
we have four first order differential equations for four variables $a$, $H$, 
$\varphi$ and $\dot\varphi$, and there is one constraint (\ref{eqn:bg1}).

As long as we consider the spatially flat universe, the system has two degrees 
of freedom for initial conditions since the scale factor is unimportant except 
its relative change.
Furthermore, because we are only interested in the region $\varphi \sim 0$ 
where the energy condition is violated and the initial singularity is 
avoided\cite{rt94}, the initial conditions actually have only one degree of 
freedom.
Since it is shown in \cite{rt94} that $\dot\varphi$ and $H$ does not 
change their signs during their evolution, we choose $\dot\varphi$ and $H$ 
to be positive in order to describe the expanding universe.
The action (\ref{eqn:rt94act}) is symmetric under the change of $\varphi$'s 
sign so that $\dot\varphi$ can be set to be positive without a loss 
of generality. 
We also assume $\lambda$ to be positive, since otherwise the system has
no singularity-free solutions\cite{art94,rt94}.
 
Homogeneous, isotropic and spatially flat solutions leading to Friedmann-like 
universe (decelerating expansion) in the future are shown in fig.1.
Since $\dot\varphi$ is always positive throughout the evolution of the 
solutions, larger value of $\varphi$ means later in time, i.e. time flows
from left to right.
We here put $\lambda=1$, which fixes the time scale. The equations of 
motion are numerically integrated from the future to the past. 
In fig.1, we fix the initial (future) value of $\varphi $ and take
several different initial values of $H$.
The initial values of $\dot\varphi$ are determined by $H$ and $\varphi$ through
the constraint (\ref{eqn:bg1}).
We see that there are 2 classes of solutions: singular and non-singular. 
The singular 
solutions (a and b in fig. 1) lead to an initial singularity as is the case 
for the usual big-bang universe.
The non-singular solutions (c, d and e) are free from  the initial 
singularity, and approach a flat space in the infinite past.
It is shown in \cite{rt94} that all solutions that inhabit in the $\varphi<0$ 
$H>0$ quarter plane regularly continue to $\varphi >0$ $H>0$ quarter plane.

Asymptotic solution of the system  can be studied
by assuming a power-law behavior for $H$ and $\dot\varphi$ as  
\begin{eqnarray}
H&\sim&\omega_1 |t|^{\beta},\\
\varphi&\sim&\varphi_0+\omega_2\ln |t|,\\
\dot\varphi&\sim&\frac{\omega_2}{t},
\end{eqnarray}
in the asymptotic region $t\rightarrow\pm\infty$\cite{art94}. 
Using equations (\ref{eqn:bg1})(\ref{eqn:bg3}) and the asymptotic form of the 
function $\xi$
\begin{equation}
\frac{\partial\xi}{\partial\varphi}\sim \mbox{sign}(\varphi)\frac\pi 3 
e^{|\varphi|},
\end{equation}
we have following asymptotic solutions for $t\rightarrow\infty$ and 
$t\rightarrow -\infty$ regions respectively:
\begin{eqnarray}
A_\infty&:&\beta = -1, \omega_1 = \frac 13, \omega_2 = \sqrt{\frac 23}
\label{eqn:apinf},\\
A_{-\infty}&:&\beta = -2, \omega_1 >0, \omega_2 = -5
\label{eqn:aminf}.
\end{eqnarray}
The future solution $A_\infty$ is obtained by balancing the terms 
which do not contain Gauss-Bonnet contribution, and therefore describe the 
situation where the 1-loop effect is negligible. On the contrary, 
$A_{-\infty}$ corresponds to the case in which the Gauss-Bonnet contribution
is dominant, and requires $\lambda>0$, which is acceptable for our non-singular
solution. 
These asymptotic solutions are uniquely determined under the conditions 
mentioned above, that is,  
the solutions are singularity-free and $H > 0$, $\dot\varphi > 0$.
To sum it up, the non-singular universe we are considering here begins with 
an asymptotically flat space at the past infinity,
and experiences super-inflation until $\varphi$ crosses zero, and then 
enters into a stage of 
ordinary Friedmann-like expansion where Gauss-Bonnet effect is negligible. 

For the analysis of perturbation we now look into a typical non-singular 
solution more closely.
To characterize the equations of state of the system
we introduce a new variables $\Gamma$ defined by
\begin{equation}
p_{\mbox{eff}}=(\Gamma-1)\rho_{\mbox{eff}}.
\label{eqn:gamma}
\end{equation}
Effective energy density $\rho_{\mbox{eff}}$ and effective pressure 
$p_{\mbox{eff}}$ used here are defined by the components
of Einstein tensor as $
\rho_{\mbox{eff}} = -G^0{}_0$ and $
p_{\mbox{eff}} = \frac 13 G^i{}_i$, respectively. Using the background
variables we can express $\Gamma$ in a more explicit form:
\begin{equation}
\Gamma=-\frac{2\dot H}{3H^2}
\label{eqn:gammaform}.
\end{equation}
Evolutions of $H$, $\varphi$ and $\Gamma$ in a typical non-singular
solution are shown in the fig.2. The friction 
$3H+\dot\alpha/\alpha$ is for later 
discussion. In the far future $\Gamma$ approaches 2,
which indicates that
the equation of state for this system in later time is nearly that 
of stiff matter, i.e. free scalar field. 
On the other hand, $\Gamma$ takes large negative values
in the far past. This  peculiar behavior of $\Gamma $ 
has a close relation to the non-singular nature of the solution.
By means of this $\Gamma$, weak and strong energy conditions 
$\rho_{\mbox{eff}}+p_{\mbox{eff}}\geq 0$, 
$\rho_{\mbox{eff}}+3p_{\mbox{eff}}\geq 0$
can be written as
$\Gamma\geq 0$,
$\Gamma\geq 2/3$
, respectively.
Thus, as can be seen from the fig.2, 
violation of those energy conditions, which is necessary to avoid
the initial singularity, occurs near and before the peak of the 
Hubble parameter $H$ at $\varphi \sim 0$.


We now consider the perturbation of our model
(\ref{eqn:rt94act}) to
analyze the stability of the non-singular solutions. We include only the tensor
perturbation here\footnote{We have also analyzed scalar and vector part of the 
perturbation using the gauge-invariant method\cite{kss98}, but no instability 
was found.}, and write the full metric as
\begin{equation}
ds^2=-dt^2+a^2(\delta_{ij}+h_{ij})dx^i dy^j
\label{eqn:tmetpert}.
\end{equation}
For $h_{ij}$ we assume the transverse-traceless conditions 
$h^i{}_i=0$, $h_{ij}{}^{|j}=0$.
Equations of motion for $h_{ij}$ are obtained by perturbing the
Einstein-like equation (\ref{eqn:Einstein}), and non-trivial one comes from
the spatial part.
Expanding $h_{ij}$ with transverse-traceless basis tensors as
\begin{equation}
h_{ij} = h_+ (k,t)\mbox{\bf e}_+{}_{ij}(k)
+h_\times(k,t)\mbox{\bf e}_\times{}_{ij}(k),
\end{equation}
we obtain the same equation for each polarization mode
\begin{equation}
\ddot h +(3H+\frac{\dot\alpha}{\alpha})\dot h +\frac{k^2}{a^2\alpha}
(1-\frac\lambda 2\ddot\xi)h =0,
\label{eqn:tenp}
\end{equation}
where $\alpha$ is a function of the background variables defined by
\begin{equation}
\alpha:=1-\frac\lambda 2 H\dot\xi=\frac{\dot\varphi^2}{6H^2}
\label{eqn:alpha}.
\end{equation}
Equation (\ref{eqn:tenp}) describes the evolution of the geometrical 
fluctuation in our non-singular model. 
Of course, this  
is nothing but the equation for the gravitational wave in FRW background
if $\lambda=0$.
The background equation (\ref{eqn:bg2}) and the definition of $\Gamma$
(\ref{eqn:gamma}) lead to
 $1-\frac\lambda 2 \ddot\xi =(3\Gamma-5)\alpha$ 
so that the perturbation equation (\ref{eqn:tenp}) can be expressed as
\begin{equation}
\ddot h +(3H+\frac{\dot\alpha}{\alpha})\dot h +
\frac{k^2}{a^2}(3\Gamma-5)h=0
\label{eqn:tenpwg},
\end{equation}
which is suitable to discuss the stability of the system in relation to
the energy conditions.
In this expression the third term determines the stability of the system, 
and the second term is regarded as a friction. One can see that the system is 
unstable for tensor perturbation when 
\begin{equation}
\Gamma<\frac 53,
\label{eqn:stcond}
\end{equation}
and that strong instability appears in small scale. 
We can combine this result with the discussion about the energy conditions
to reach the statement, that is, {\it Cosmological solution of this
model is geometrically unstable as long as it is non-singular.}
In order to avoid the initial singularity, the strong energy condition
$\Gamma\geq 2/3$ must be violated. Then, since $\Gamma<2/3<5/3$, condition
(\ref{eqn:stcond}) is satisfied and the system becomes unstable for tensor
perturbation.
Therefore, the non-singular nature of the background space-time inevitably 
requires its geometrical instability, which is characterized by 
the condition (\ref{eqn:stcond}).

Numerical solutions of perturbation for different wave numbers 
in the flat background are shown in the fig.3. 
The ordinate is in logarithmic scale,
and we can easily recognize that the instability arises 
in the super-inflationary phase. In these 
calculations the initial value of $h$ is set to unity so that  
$log_{10}h$ starts from zero. Our statement  can be confirmed by 
comparing fig.2 
and fig.3. One interesting point we can notice is that the effect of the
friction term $3H+\dot\alpha/\alpha$ is not necessarily negligible.
The friction term is negative before the  peak of the Hubble parameter 
and this negative friction
enhances the instability. After the peak, the friction turns
to be positive and it weakens the growth of the perturbation. This is why the
instability seems to cease before $\Gamma$ gets greater than $5/3$.

One might wonder whether this instability continues from the infinite past or 
not. To answer this question we can use the asymptotic background solution
$A_{-\infty}$ which tells us the asymptotic form of $\Gamma$ 
in the infinite past.
Substituting (\ref{eqn:aminf}) into (\ref{eqn:gammaform}), we have the 
asymptotic form of $\Gamma$ as $\Gamma\sim 4t/3\omega_1$. Since 
$\omega_1$ is positive, $\Gamma$ decreases indefinitely in the
past and the energy conditions can never be recovered. From the condition
(\ref{eqn:stcond}), 
therefore, the geometrical instability never disappears even if we go to the
infinite past. To obtain the asymptotic solution of $h$ in 
the past asymptotic region, we estimate each term of the 
equation(\ref{eqn:tenpwg}). As $\dot\alpha/\alpha$ behaves like $1/|t|$ and
$H\sim\omega_1/|t|^2$ in the far past, the friction term in
(\ref{eqn:tenpwg}) can be neglected. In the third term, the scale factor $a$ behaves like
$a=\exp{\int H dt}\sim \exp(\omega_0 -\frac{\omega_1}{t})\sim \exp{\omega_0}
=:a_0$ ($\omega_0$ and $a_0$ are constants). Therefore, (\ref{eqn:tenpwg}) 
becomes
\begin{equation}
\ddot h+\frac{4k^2}{\omega_1a_0{}^2}th=0.
\end{equation}
and the asymptotic solution of $h$ can be expressed using Airy functions
as $h(k,t)\sim C_1 Ai(\tau)+C_2 Bi(\tau)$, where 
$\tau = \sqrt[3]{\frac{4k^2}{\omega_1a_0{}^2}}|t|$. 

The fact that the perturbative instability continues from the infinite
past does not necessarily mean the failure of the background solution.
To see the criterion for the breakdown of the background solution, 
we assume a state of the perturbation at some finite past time 
$t=t_i$ as $h(k,t_i)=h_i(k)$ and $h'(k,t_i)=h'_i(k)$ 
(prime denotes differentiation with respect to $\tau$).
Then, $C_1$ and $C_2$ are determined and we can write
\begin{equation}
h(k,t)=\pi\left(Bi'(\tau_i)h_i(k)-Bi(\tau_i)h'_i(k)\right)Ai(\tau)
      -\pi\left(Ai'(\tau_i)h_i(k)-Ai(\tau_i)h'_i(k)\right)Bi(\tau),
\end{equation}
where $\tau_i=\sqrt[3]{\frac{4k^2}{\omega_1a_0{}^2}}|t_i|$. 
In order to keep the perturbation smaller than the back ground metric through
the evolution of the solution, 
$h(k,t)$ has to be smaller than unity when $t\sim0$, which is sufficient since 
the growth of the perturbation terminates around the Hubble peak.
Then, discarding the decaying mode and picking up the dominant term 
($h_i(k)$ and$h'_i(k)$ are assumed to be comparable in magnitude), 
the background solution breaks down if and only if
$h(k,0)= \pi Bi'(\tau_i)h_i(k)Ai(0)\gg{\cal O}(1)$. 
We can use the asymptotic form of the Airy function to put this breaking down 
condition expressed as a condition for the value of the perturbation at time 
$t_i$
\begin{equation}
h_i(k)\gg\pi^{-\frac{1}{2}}Ai(0)^{-1}
\left(\frac{4k^2}{\omega_1a_0{}^2}\right)^{-\frac{1}{12}}|t_i|^{-\frac 14}
\exp\left(-\frac{4|k|}{3a_0\omega_1}|t_i|^{\frac 32}\right),
\label{eqn:pertcond}
\end{equation}
where $Ai(0)^{-1}\approx 2.817 \sim{\cal O}(1)$.
Therefore, premising the existence of some natural cut-off for $k$, we can say
that the background system breaks down due to the overdominant graviton if the
amplitude $h_i(k)$ of at least one mode $k$ satisfies the condition 
(\ref{eqn:pertcond}) at time $t_i$.


We have shown that the non-singular cosmological model proposed by Antoniadis, 
Rizos and Tamvakis\cite{art94} is unstable for tensor perturbation, and that 
the 
energy condition breaking required to obtain the non-singular solutions is a
sufficient condition for this instability. We also estimated the condition for 
the break down of the background solution due to the overdominant
graviton(\ref{eqn:pertcond}).

Let us conclude this letter by giving some comments on our result.
First, we would like to comment on the nature of the instability found here.
One can say that this instability has an opposite nature to Jeans
instability, in the sense that it appears in small scales rather than in large
scales. Apart from its characteristic spectrum, this instability is distinctive
in that it appears only in the tensor part. Since it involves no growth of 
scalar field perturbation\cite{kss98}, this instability is purely 
a ``geometrical'' one.
If we assume that the effective action (\ref{eqn:ss1leAct}) is valid in the 
early stages of our universe, and if none of the $h_i(k)$ satisfies the 
condition (\ref{eqn:pertcond}) at time $t_i$ so that
the background solution is unaffected by the perturbation, 
this small scale instability may indicate generation of many small
primordial black holes, typically of the Planckian scale.
Next, although we have treated only the spatially flat case so far, 
our claim also holds for the metric with non-zero spatial curvature
\footnote{In that case, 
$\Gamma=-\frac 23 \frac{\dot H-{\cal K}/a^2}{H^2+{\cal K}/a^2}$ 
and $k^2$ in (\ref{eqn:tenpwg}) is replaced by $k^2+2{\cal K}$,
where ${\cal K}=+1, -1$ for closed and open universes, respectively.}.
However, if the background universe has positive or negative spatial curvature,
the asymptotic behavior of the system in the infinite past is not the same as 
in the flat case\cite{maeda96}. 
Therefore, if the energy conditions should be recovered in the finite past, 
the instability might no longer be eternal. 
Lastly, we have to mention the possibility that the effective action we use
here (\ref{eqn:ss1leAct}) might not be suitable for the model of the early 
universe. 
Since the 1-loop correction changes the tree-level solutions drastically, 
it is sensible to consider that the higher loop effect might change 
the nature of the cosmological solutions from that of 1-loop action. 
Our results of rather pathological feature of the instability might be a
consequence of our neglecting higher corrections.
To take all these loop effects into account, we have to wait for the 
accomplishment of a non-perturbative string theory.
\acknowledgements
 
This work of M.S. is supported in part by Grant-in-Aid for Scientific 
Research 09226220.

\newpage
{\LARGE Figure captions}\\

Fig.1\\
 Flat(${\cal K}=0$) background solutions that lead to decelerating expansion 
in the future. Numerical integration is performed from future to past,
and the figure shows the phase diagram of Hubble parameter $H$ versus modulus
field $\varphi$. Time flows from left to right since $\dot\varphi$ is always
positive.
There are two classes of solutions: singular solutions (a and b), 
and non-singular solutions (c,d,e).  
One can see that the avoidance of the initial singularity takes place when
$\varphi\sim 0 $.
For detailed discussions see references \cite{art94} and  \cite{rt94}.\\

Fig.2\\
 Evolutions of (i)$H$, (ii)$\varphi$, (iii)$3H+\dot\alpha/\alpha$ and (iv)
$\Gamma$ in a typical super-inflationary flat solution. 
The origin of time $t$ is chosen so that $\varphi$ crosses zero when $t=0$. 
$H$ has a peak at $t\sim 0$. 
$\varphi$ accelerates from the asymptotically flat space, and decelerates at 
$\varphi\sim 0 $.
The friction term $3H+\dot\alpha/\alpha$
in the perturbation equation of motion 
evolves from a small negative value, 
increases in its magnitude, and changes its sign at $t=0$, and then 
decreases in the future. 
$\Gamma$ varies from a large negative value in the past to $2$ in the 
future. \\

Fig.3\\
 Evolution of the tensor perturbation in a flat background. Note that the 
tensor perturbation is plotted with logarithmic scale. We set $a=1$, $h(k)=1$
at the
onset of the integration ($t=-2$), and $t=0$ is the time
when $\varphi$ crosses zero. One can notice that growing modes appear
in the initial super-inflationary phase. The smaller the scale of perturbation
is, the larger the growth rate becomes. The depression of $k=20$ solution
around $t\sim 1.25$ indicates the oscillatory behavior in the Friedmann-like
phase. 

\newpage

\input epsf.sty
\pagestyle{empty}
\begin{figure}
\epsfxsize=16cm
\epsfysize=12cm
\epsffile{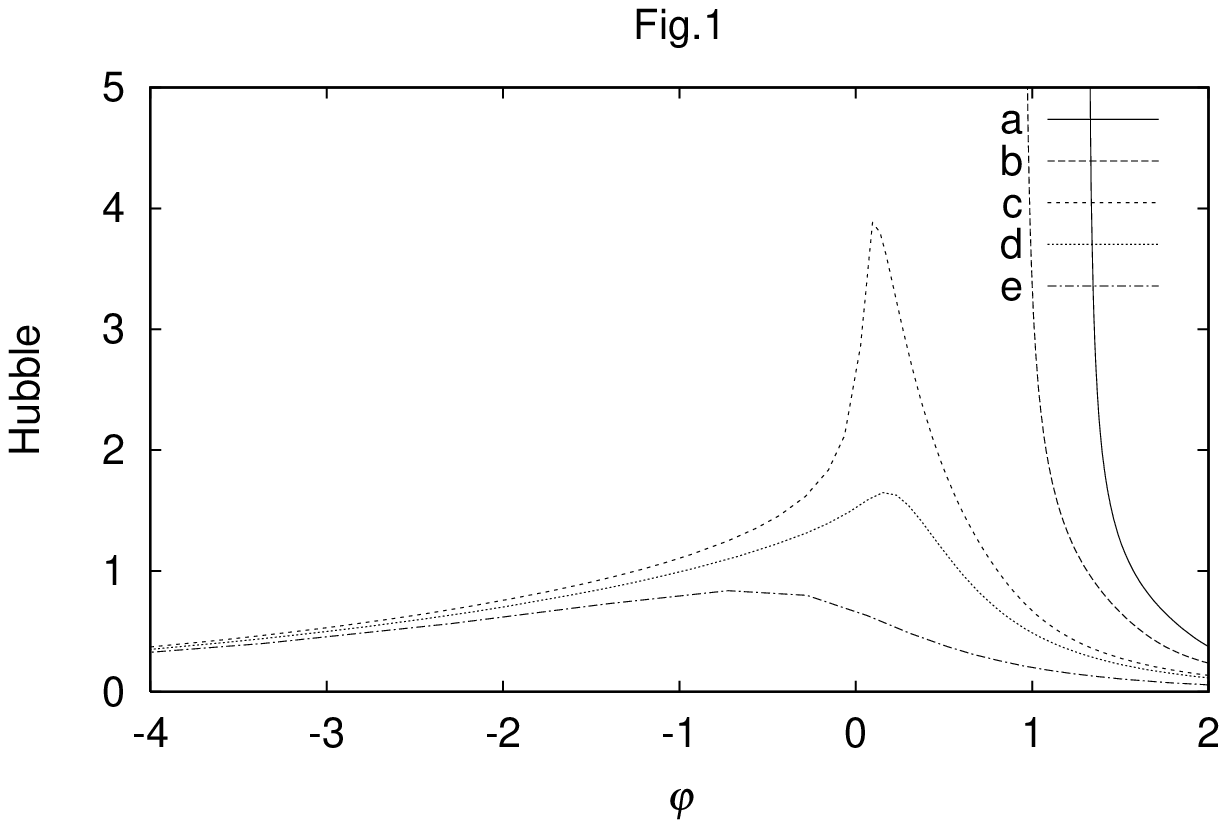}
\end{figure}
\newpage
\begin{figure}
\epsfxsize=16cm
\epsfysize=5cm
\epsffile{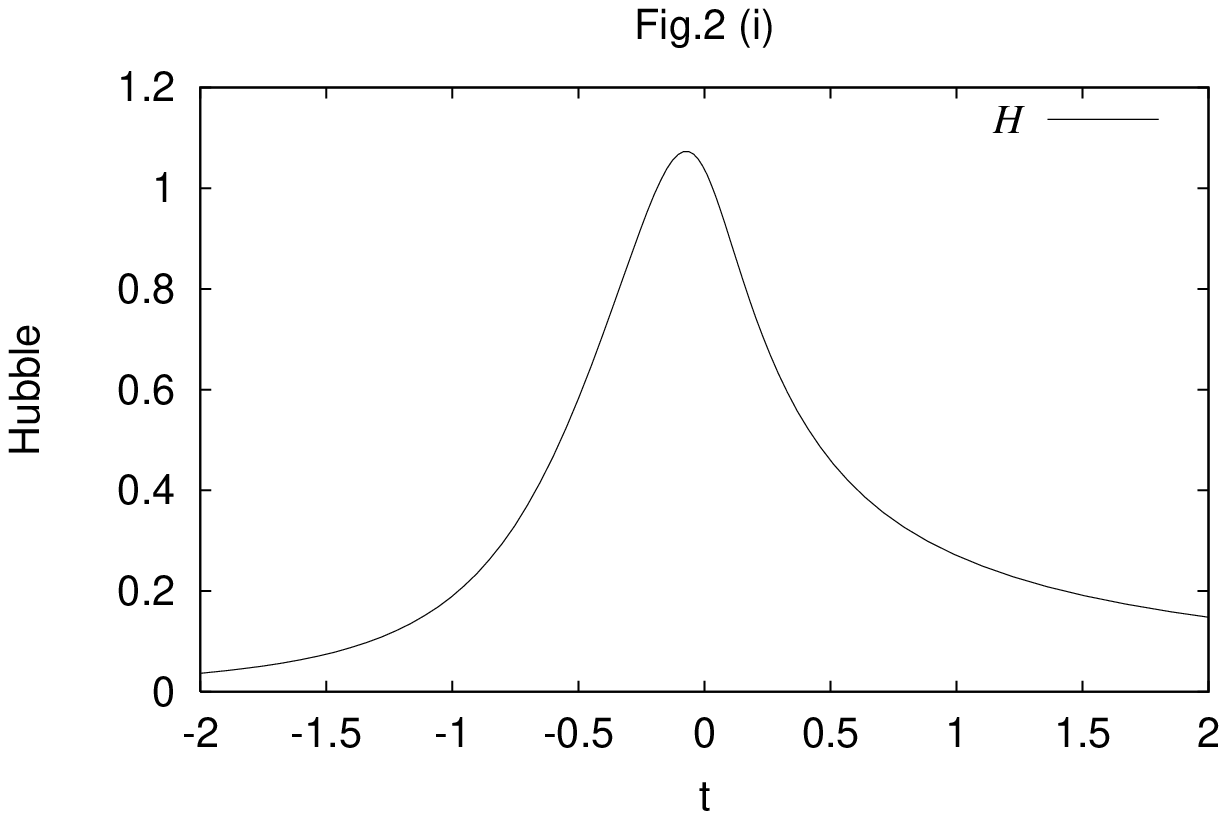}
\end{figure}
\begin{figure}
\epsfxsize=16cm
\epsfysize=5cm
\epsffile{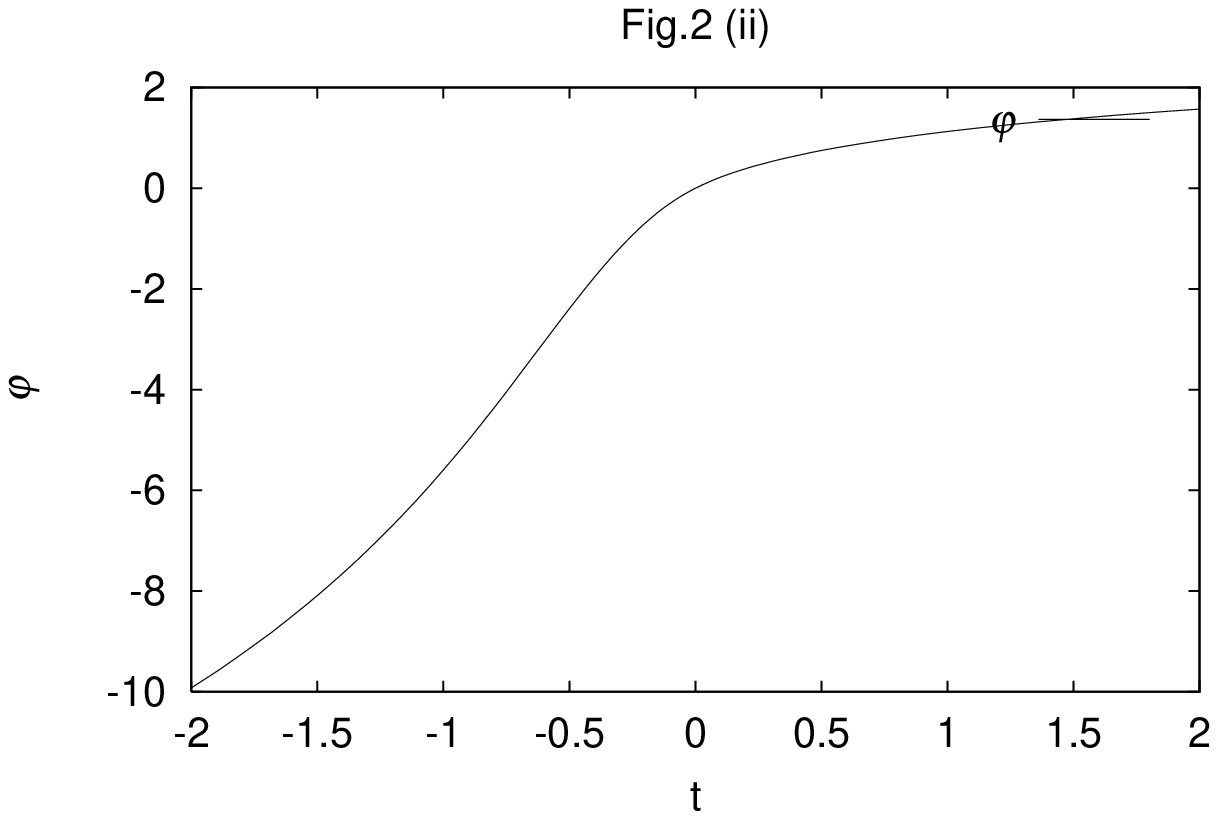}
\end{figure}
\begin{figure}
\epsfxsize=16cm
\epsfysize=5cm
\epsffile{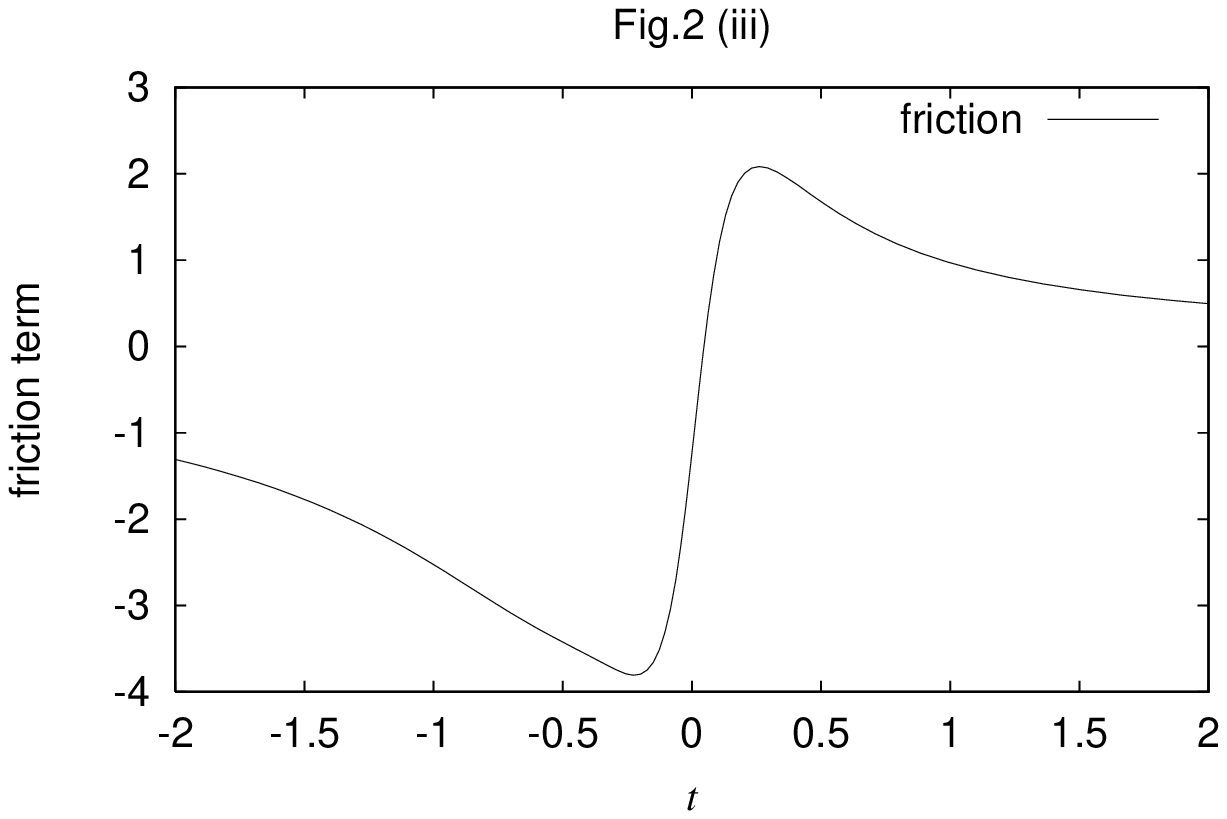}
\end{figure}
\begin{figure}
\epsfxsize=16cm
\epsfysize=5cm
\epsffile{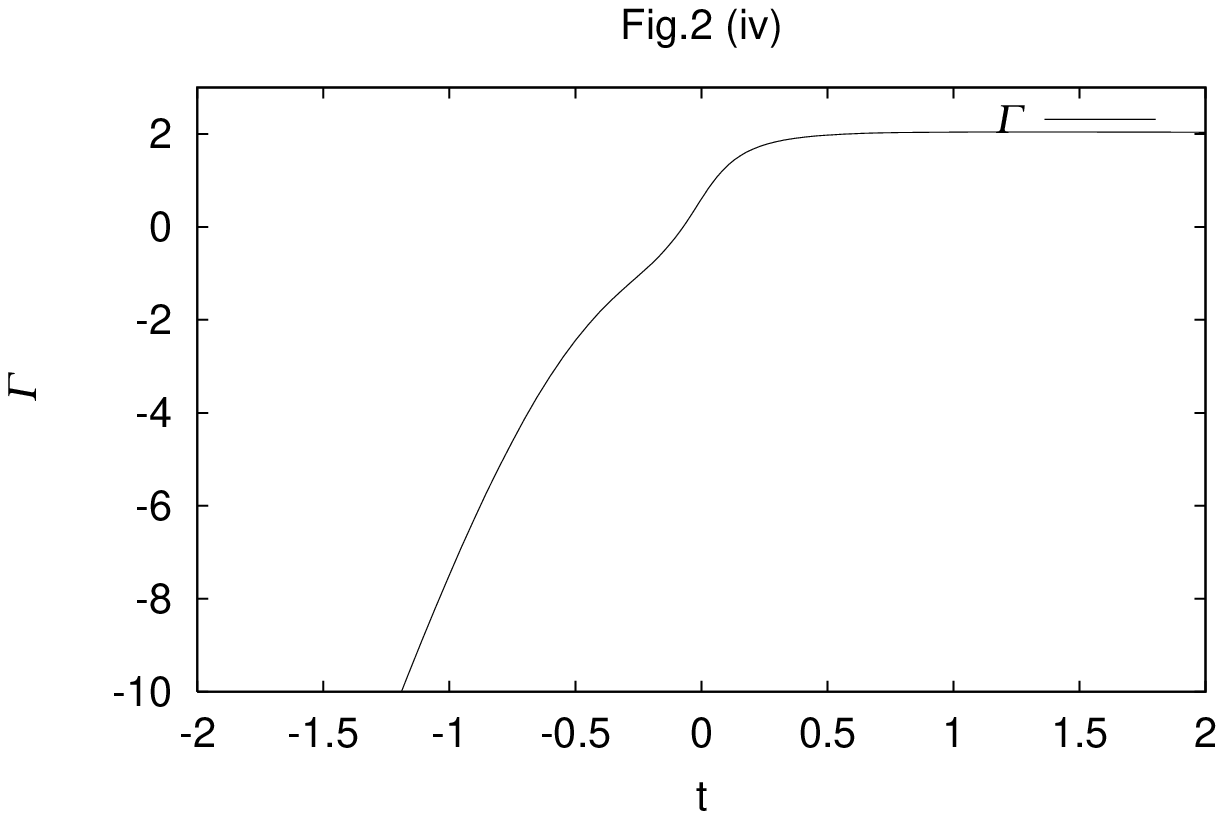}
\end{figure}
\newpage
\begin{figure}
\epsfxsize=16cm
\epsfysize=12cm
\epsffile{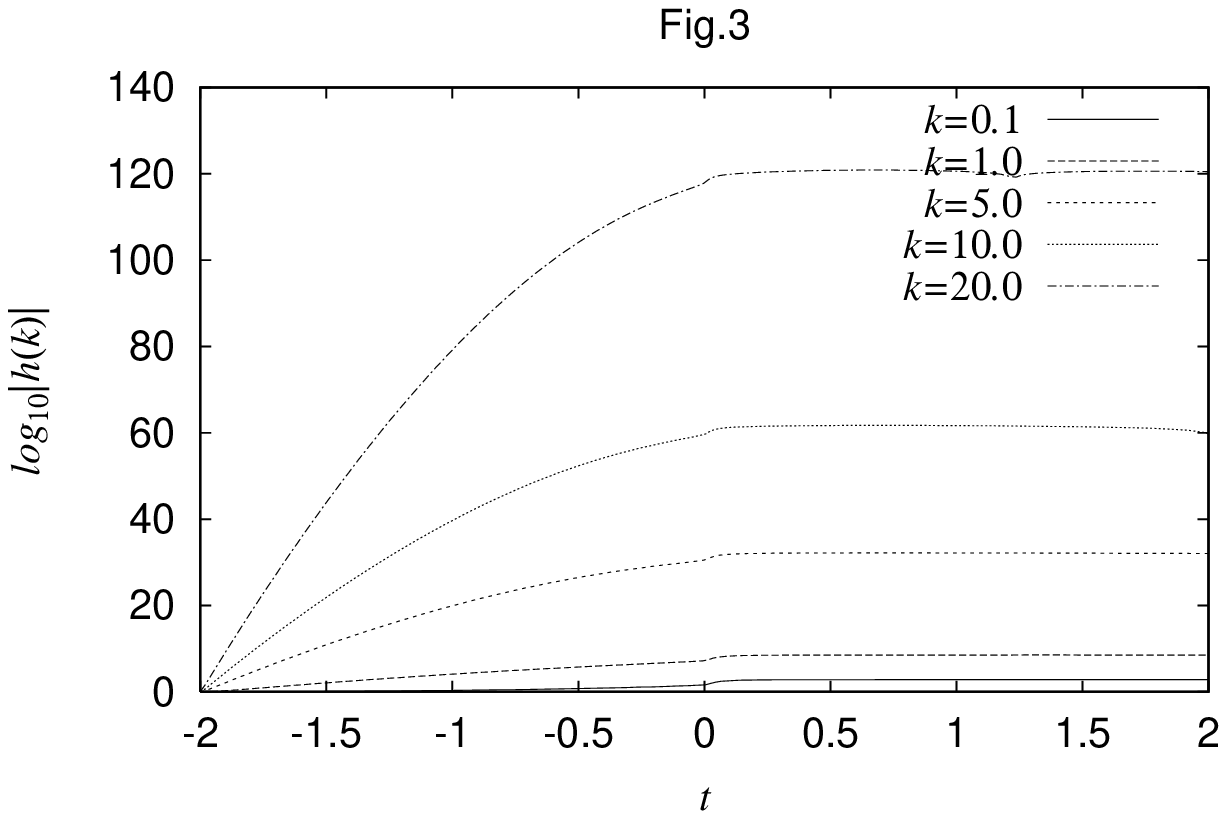}
\end{figure}

\end{document}